\documentclass[twocolumn,english]{IEEEtran}
\usepackage[T1]{fontenc}
\usepackage{babel}
\usepackage{float}
\usepackage{amsmath}
\usepackage{graphicx}
\usepackage{setspace}
\usepackage[unicode=true,
 bookmarks=true,bookmarksnumbered=true,bookmarksopen=true,bookmarksopenlevel=1,
 breaklinks=false,pdfborder={0 0 0},pdfborderstyle={},backref=false,colorlinks=false]
 {hyperref}
\hypersetup{pdftitle={Your Title},
 pdfauthor={Your Name},
 pdfpagelayout=OneColumn, pdfnewwindow=true, pdfstartview=XYZ, plainpages=false}

\makeatletter
\ifCLASSOPTIONcompsoc
\usepackage[caption=false,font=normalsize,labelfont=sf,textfont=sf]{subfig}
\else
\usepackage[caption=false,font=footnotesize]{subfig}
\fi
\usepackage{setspace}
\@ifundefined{showcaptionsetup}{}{%
 \PassOptionsToPackage{caption=false}{subfig}}
\usepackage{subfig}
\makeatother

\begin{document}
\title{An N-Path Filter with Multiphase PWM Clocks for Harmonic Response
Suppression}
\author{Venkata S. Rayudu, Heechai Kang and R. Gharpurey\\
The University of Texas at Austin}
\maketitle
\begin{abstract}
A switched-capacitor $N$-path circuit can be employed for filtering
an RF signal, as well as a passive downconverter. A known limitation
of an \textit{N}-path filter is that in addition to downconverting
signals around the desired center frequency, the circuit also downconverts
signals located around harmonics of the center frequency. An \textit{N}-path
filter that uses a PWM representation of a sinusoidal LO to mitigate
harmonic downconversion is proposed in this work. Single-edge natural-sampling
pulse-width modulated (PWM) clocks are used to drive the switches
in the \textit{N}-path filter. The potential for employing PWM for
providing gain control is also described. \bstctlcite{IEEEexample:BSTcontrol}
\end{abstract}

\begin{IEEEkeywords}
Pulse-Width Modulation (PWM), N-path filter, Harmonic Response \thispagestyle{empty}
\end{IEEEkeywords}

\section{Introduction\label{sec:Introduction}}

\begin{figure}[tbh]
\centering{\includegraphics[width=0.65\columnwidth]{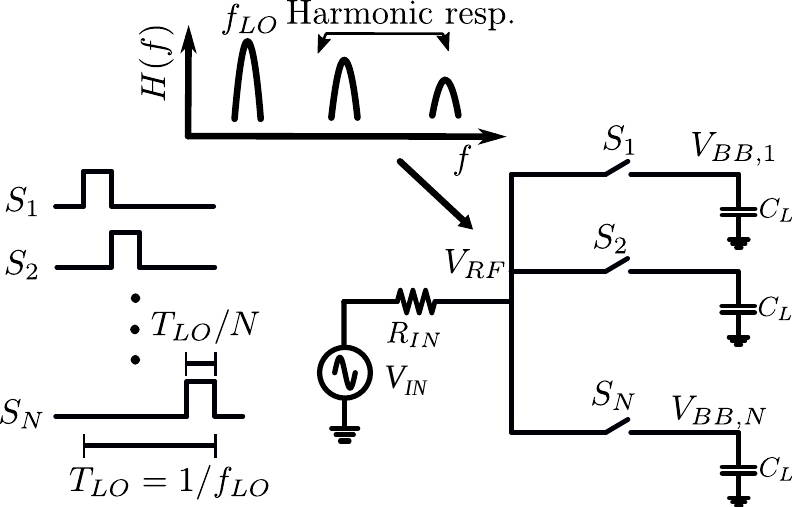}}
\centering{}\caption{N-path filter and fundamental response at $f_{LO}$\label{fig:SimplifiedN-path}}
\end{figure}

N-path filters frequency-translate the response of baseband filters
to around an LO, thereby providing the frequency selectivity of a
baseband circuit at a higher frequency \cite{Franks1960}. Switch-based
implementations of such filters have been demonstrated in recent literature
e.g., \cite{Cook2006}\cite{Ghaffari2010}\cite{Duipmans2015}\cite{Han2019}.
\textit{N}-path filters (Fig. \ref{fig:SimplifiedN-path}) can provide
a high effective quality factor, a high dynamic range and a tunable
response, in addition to being scalable with process technology \cite{Darvishi2013}.
The filters allow for relaxing linearity at their input by attenuating
blockers.

A typical CMOS implementation of a switch-based \textit{N}-path filter
employs non-overlapping pulse-based clocks of equal duty cycle of
(1/$Nf_{LO}$), where $f_{LO}$ is the frequency to which the baseband
filter response is translated, and $N$ is the number of paths in
the \textit{N}-path filter (Fig. \ref{fig:SimplifiedN-path}). For
capacitive baseband loads ($C_{L}$ in Fig. \ref{fig:SimplifiedN-path}),
an input signal at $f_{LO}\pm f_{in}$ is effectively filtered by
a bandpass response whose bandwidth is determined by $R_{IN}$, $C_{L}$
and $N$. The \textit{N}-Path filter also frequency translates an
input signal at $f_{LO}\pm f_{in}$ to a baseband signal at $f_{in}$
(nodes $V_{BB,n}$ in Fig. \ref{fig:SimplifiedN-path}).

With $N$ switching paths, input signals around $(k\times N\pm1)f_{LO}$
are translated to around $f_{LO}$ at the input node, which is termed\textit{
harmonic-folding. }This problem can be alleviated by increasing $N$,
which effectively pushes out the frequencies from where harmonic folding
can occur farther out in frequency, compared to the desired signal
input at $f_{LO}\pm f_{in}$. If combined with a band-limiting prefilter,
the high-frequency components can be attenuated before they can fold
in-band.

Another design challenge is that of \textit{harmonic response}. In
addition to $f_{LO}$, the baseband response is also translated to
$kf_{LO}$ and conversely, an input signal located at $kf_{LO}\pm f_{in}$,
is translated to baseband, where for differential implementations,
$k$ is odd. This can degrade performance in the presence of an interferer
around $kf_{LO}$, since the interferer after downconversion to baseband
would overlap with the signal. An approach to reduce harmonic responses
in an $N$-path filter was shown in \cite{xu18}, which combined a
harmonic-rejection front-end employing transconductor segments with
a $1:\sqrt{2}:1$ gain ratio with $N$-path filters to suppress harmonic
responses. $N$-path filtering was implemented at the transconductor
outputs. The filter does not relax linearity at the input of the transconductor. 

Ideally, if sinusoidal LOs are used for frequency-translation, as
theoretically analyzed in \cite{Franks1960}, the problem of harmonic
response is avoided. However, a CMOS switch-based approach does not
allow for directly employing a sinusoidal LO.

On the other hand, by employing PWM, a sinusoid can be represented
by an equivalent switching waveform, which was demonstrated for providing
harmonic rejection in a mixer in \cite{hkang18}. PWM is a signaling
scheme, wherein the amplitude information of an analog signal is represented
by the duty-cycle of a periodic discrete-level pulse train, that has
a frequency that is significantly higher than the highest frequency
of the analog signal \cite{Nielsen_PWM}. Since the PWM signal is
a discrete-level waveform, it can be easily combined with a switch-based
$N$--path filter, without the requirement for linear multipliers.

In this work, a PWM-based clocking scheme to reduce harmonic responses
in a passive $N$-path filter is proposed. The architecture, including
clock-shaping are described in Section II. The approach also allows
for gain control and a tunable input impedance match. Section III
describes simulation-based results and the conclusion follows in Section
IV.

\section{Overview of the Architecture\label{sec:Overview-of-Architecture}}

\begin{figure}[tbh]
\centering{\includegraphics[width=0.8\columnwidth]{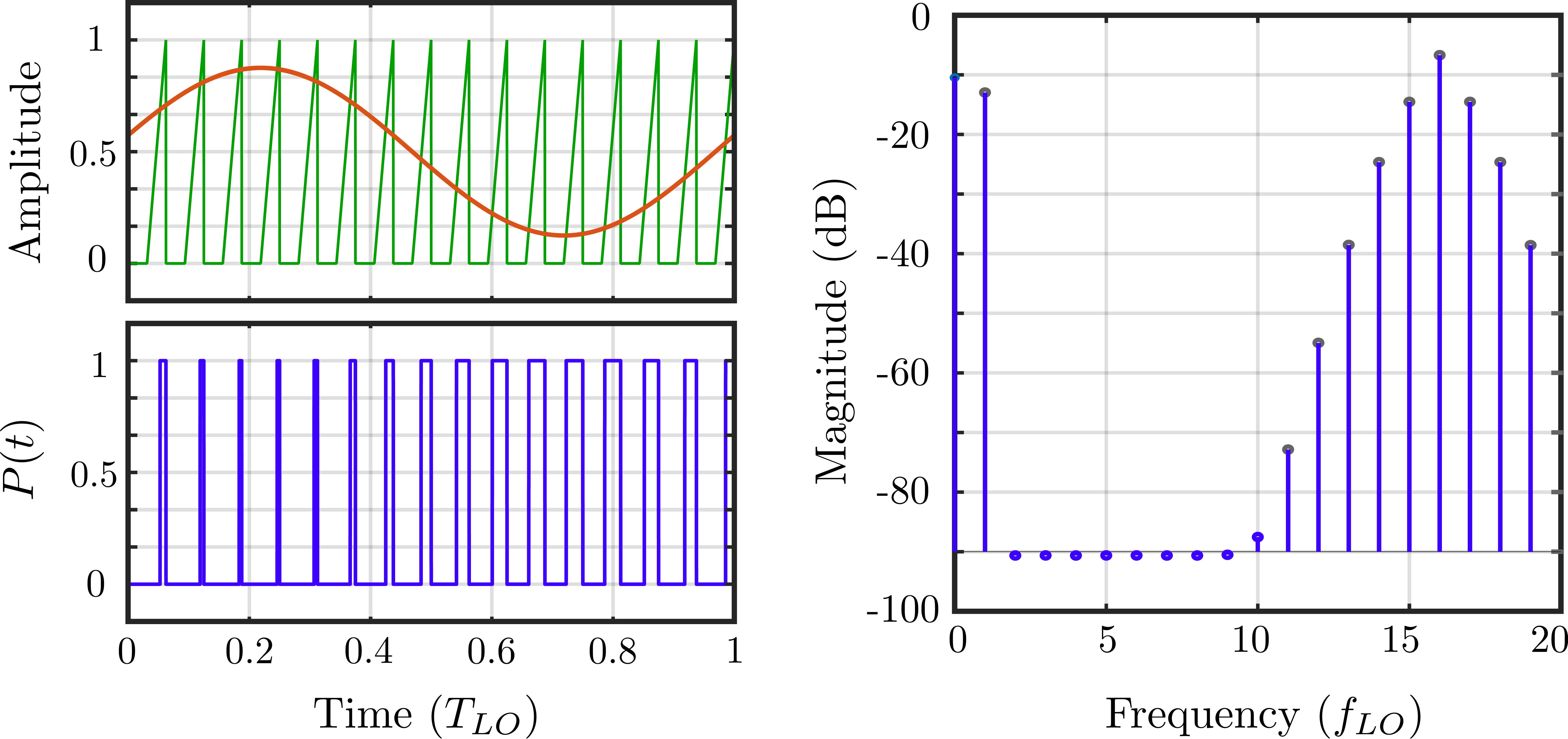}

\caption{Trailing-edge PWM. (a) Waveform. (b) Spectrum \label{fig:Trailing edge PWM-1}}
}
\end{figure}

A PWM signal corresponding to a sinusoid and its spectrum are shown
in Fig. \ref{fig:Trailing edge PWM-1}. The PWM signal is generated
by taking the difference of a continuous-time sinusoid of frequency
$f_{LO}$, from a single-edge ramp waveform, and hard-limiting the
difference, based on its polarity. The ramp is assumed to be periodic
with a frequency of $f_{PWM}$ (period $T_{PWM}=1/f_{PWM}$), where
$f_{PWM}=16f_{LO}$. In general, $f_{PWM}\gg f_{LO}$. The closest
significant harmonics in the PWM spectrum are located far from the
$f_{LO}$ component (Fig. \ref{fig:Trailing edge PWM-1}).

The PWM signal, $P(t)$, consists of a train of pulses, whose duty
cycle varies in response to the sinusoidal LO of frequency $f_{LO}$.
The PWM pulse-train exhibits a periodicity of $T_{LO}$, where $T_{LO}=1/f_{LO}$,
and thus can be expressed as a Fourier Series in the angular frequency
$\omega_{LO}$, where $\omega_{LO}=2\pi f_{LO}$.
\begin{align}
P(t) & =\sum_{n=0}^{\infty}\left\{ a_{n}cos(n\omega_{LO}t)+b_{n}sin(n\omega_{LO}t)\right\} \label{eq:PWMFS}
\end{align}

The time-domain representation of $P(t)$ can be determined by solving
for the intersection points of the ramp waveforms with the sinusoid.
The coefficients $a_{n}$ and $b_{n}$ can then be derived from $P(t)$.
$P(t)$ in Fig. \ref{fig:Trailing edge PWM-1}, for a sinusoid amplitude
of $A_{LO},$ can be shown to have a DC component of 0.25, and a fundamental
amplitude of $0.5A_{LO}$.

The proposed architecture employs the above PWM sinusoids in the switching
paths of Fig. \ref{fig:SimplifiedN-path}. For $N=4$, four PWM LOs
that provide $f_{LO}$ components with relative phase shift of $90^{\circ}$
are required. The PWM-LOs must be non-overlapping and contiguous so
that one and only one load capacitor is connected to the input at
any time. This avoids charge-sharing between the load capacitors,
or an open-input, that can severely degrade the frequency response.

\begin{figure}[tbh]
\centering\includegraphics[width=0.8\columnwidth]{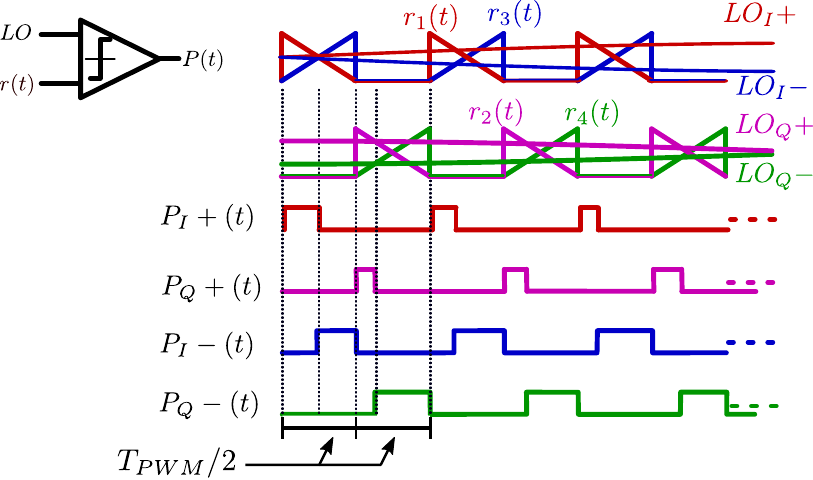}\caption{I and Q path PWM waveform generation \label{fig:PWM-waveforms}}
\end{figure}

$I$ and $Q$ PWM clocks are synthesized by comparing sine-waves of
amplitude $A_{LO}$ and frequency $f_{LO}$ with quadrature phase-offset,
in two consequent time-halves of one PWM pulse period $T_{PWM}$ (Fig.
\ref{fig:PWM-waveforms}). This ensures that the $I$ and Q PWM pulses,
$P_{I}(t)$ and $P_{Q}(t)$ are non-overlapping. Further, PWM pulses
are generated to synthesize out-of-phase $f_{LO}$ components in the
$I$ and $Q$ paths each.

A sinusoid with phase $0$ is compared to a leading-edge ramp $r_{1}(t)$,
for example, that is non-zero for $0<t<T_{PWM}/2$. This provides
the positive PWM-LO $P_{I}+(t)$. Comparing a sinusoid of phase $180^{\circ}$
to a trailing-edge ramp, $r_{3}(t)$, in the same time then provides
the complementary PWM-LO $P_{I}-(t)$. $P_{I}+(t)$ and $P_{I}-(t)$
together span a time duration of $T_{PWM}/2$. To generate the positive
$Q$ PWM-LO, $P_{Q}+(t)$, a sinusoid with phase $90^{\circ}$ is
compared to a leading-edge ramp, $r_{2}(t)$, that is non-zero for
$T_{PWM}/2<t<T_{PWM}$. The sinusoid of phase $270^{\circ}$, is compared
to trailing-edge ramp $r_{4}(t)$ that is non-zero in the same time
window, to provide $P_{Q}-(t)$. $P_{I}\pm(t)$ and $P_{Q}\pm(t)$
together span the full PWM pulse duration, $T_{PWM},$ and are non-overlapping.
Since we assume, $f_{PWM}=16f_{LO}$, 16 ramp pulses are applied to
each phase of a full cycle of an $LO$ sinusoid.

 The waveform pairs $P_{I}\pm(t)$ and $P_{Q}\pm(t)$ are not strictly
out-of-phase, even though the phases of the frequency components at
$f_{LO}$ differ by $180^{\circ},$ since they are generated by different
ramp signals (e.g, $r_{3}$ and $r_{1}$). Consequently, the baseband
signals observed on the $I+$ and $I-$, and similarly $Q+$ and $Q-$
paths, are not ideally out-of-phase.

\begin{figure}[tbh]
\centering\includegraphics[width=0.7\columnwidth]{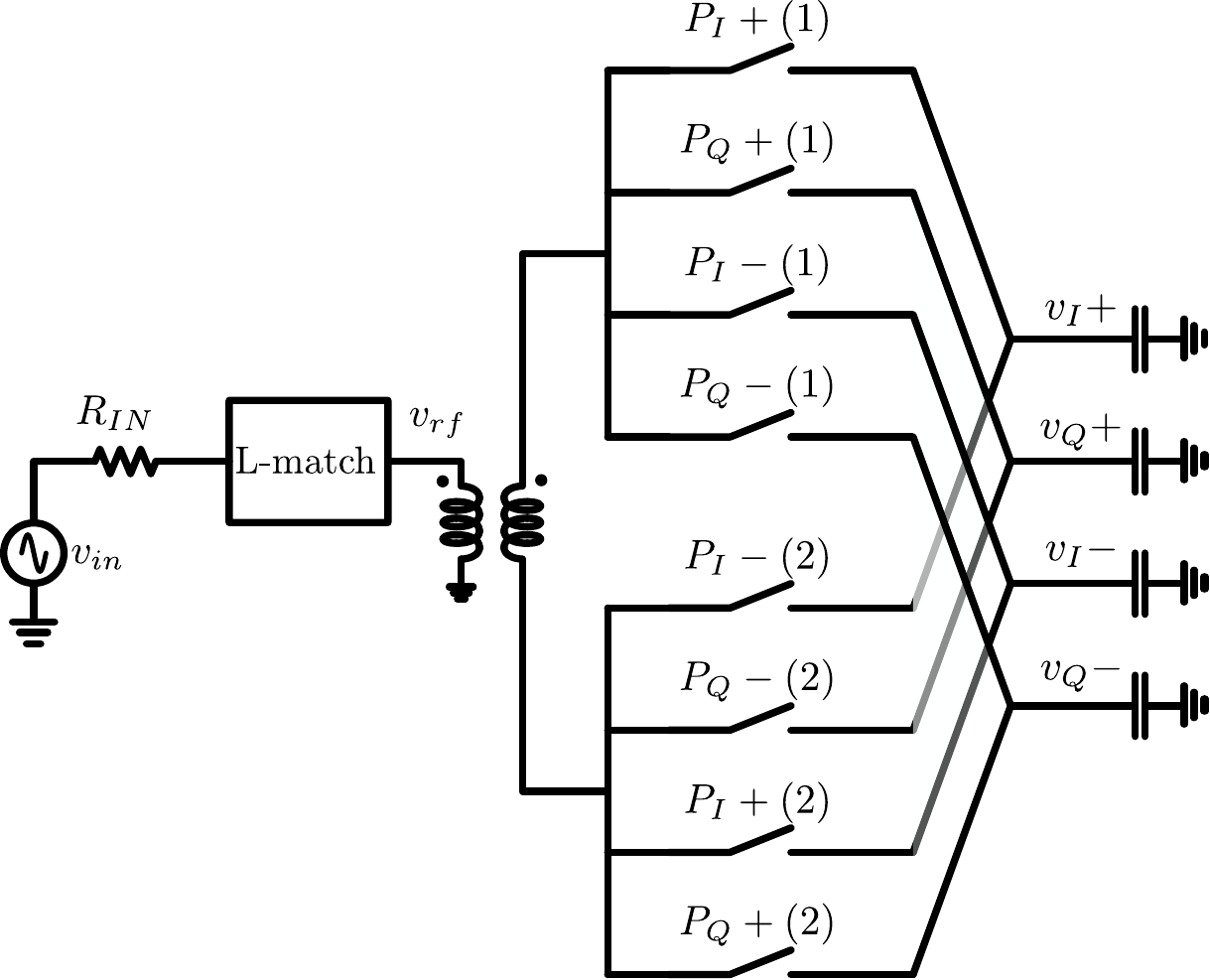}\caption{PWM-LO N-path filter architecture\label{fig:Generation-of-symmetrically}}
\end{figure}

The above issue is addressed by employing a transformer to split the
input signal into differential signal components. Each transformer
output is applied to a separate 4-path filter, and the outputs of
each filter are combined at baseband, similar to a classical $N$-path
filter (Fig. \ref{fig:Generation-of-symmetrically}). The PWM waveforms
employed in the first $N$-path filter are denoted by $P_{I,Q}\pm(1)$,
while those in the second $N$-path are given by $P_{I,Q}\pm(2)$.

The transformer provides a phase-split of 180$^{\circ}$ between its
two outputs. Thus, 4-path filter outputs with clocks of opposite polarities
need to be combined. Thus, for example, the $P_{I}+(1)$ path is combined
with the $P_{I}-(2)$ path. All harmonic components of these two paths
should be ideally out-of-phase. For this reason, both $P_{I}+(1)$
and $P_{I}-(2)$ employ the same ramp waveform in any given $T_{PWM}$
time window. The same argument is applicable for all other signal
combinations $[P_{I}-(1),P_{I}+(2)]$, $[P_{Q}+(1),P_{Q}-(2)]$ and
$[P_{Q}-(1),P_{Q}+(2)]$.

In a differential sense, e. g., $(P_{I}+(1)$-$P_{I}-(2))$, this
clocking configuration applies a $3$-level PWM switching between
levels -1, 0 and 1, to the incident signal. This lacks an $f_{PWM}$
component, which improves quadrature performance.

\begin{figure}[tbh]
\centering\includegraphics[width=0.8\columnwidth]{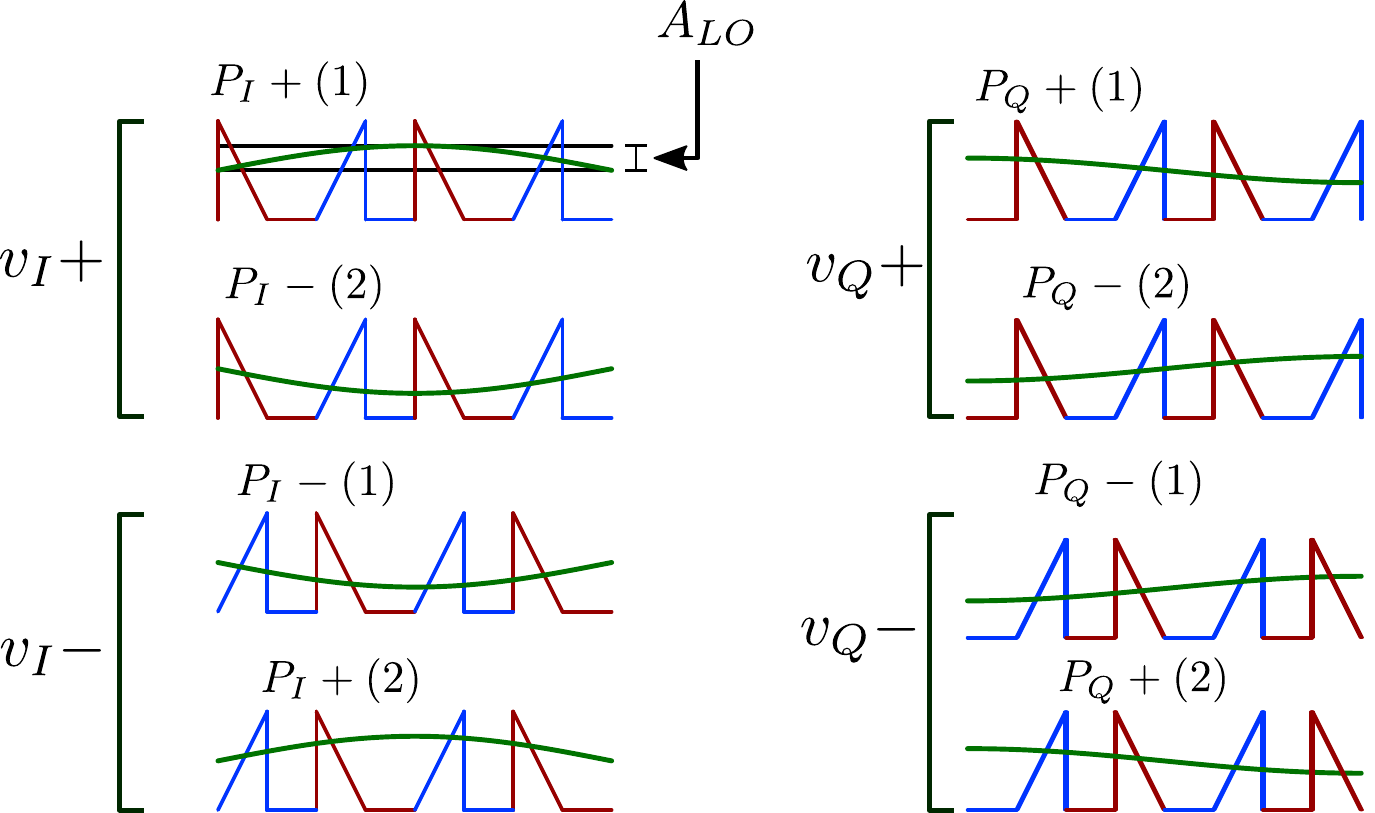}\caption{PWM-LOs with alternating ramps \label{fig:PWM-LOs-with-alternating}}
\end{figure}

It is possible to assign a single ramp to any one path of the filter
(Fig. \ref{fig:PWM-waveforms}). While the differential LO pairs $(P_{I}+(1)$-$P_{I}-(2))$
and $P_{I}-(1)-(P_{I}+(2))$, are identical waveforms with $180^{\circ}$
relative phase, the common-mode components of the LOs, namely, $(P_{I}+(1)$+$P_{I}-(2))$
and $(P_{I}+(2)$+$P_{I}-(1))$, are different. This makes the design
sensitive to capacitive parasitics at the transformer outputs. To
avoid this issue, an LO waveform shown in Fig. \ref{fig:PWM-LOs-with-alternating}
is used, which alternates the ramp applied in each path. The 3-level
differential PWM waveform is the same as the above case, but here
the common-mode of the LO pairs is identical, and its current contribution
in the presence of parasitics gets rejected as a common-mode at the
transformer input ($v_{rf}$).

While the fundamental component of waveform moves to $f_{PWM}/2$,
this component appears as a common-mode in the spectrum of the 3-level
PWM, and is rejected at the input, and significant PWM harmonics are
still located around $f_{PWM}$.

Since in each path, the frequency translation is performed by a sinusoid
instead of a rectangular waveform, the harmonic responses observed
in a classical $N$-path filter are avoided. For $f_{PWM}=16f_{LO}$,
there is no significant harmonic response at harmonics from $2f_{LO}$
to $13f_{LO}$ (Fig. \ref{fig:Trailing edge PWM-1}). An input series
inductor is included in the design \cite{Duipmans2015}.

\vspace{-0.4cm}

\subsection{\label{subsec:PWM-Pulse-Generator}PWM Pulse Generator}

In the description above, we have assumed that the PWM pulses are
generated by comparing a sinusoidal signal to a ramp in an ideal comparator.
While this is theoretically possible, for applications requiring LOs
in the tens to hundreds of MHz, this would imply the use of GHz-rate
PWM ramps, which are difficult to implement. Additionally, this would
also require comparators with gain-bandwidth products in the range
of hundreds of GHz, which is impractical.

\begin{figure}[tbh]
\centering\includegraphics[width=0.6\columnwidth]{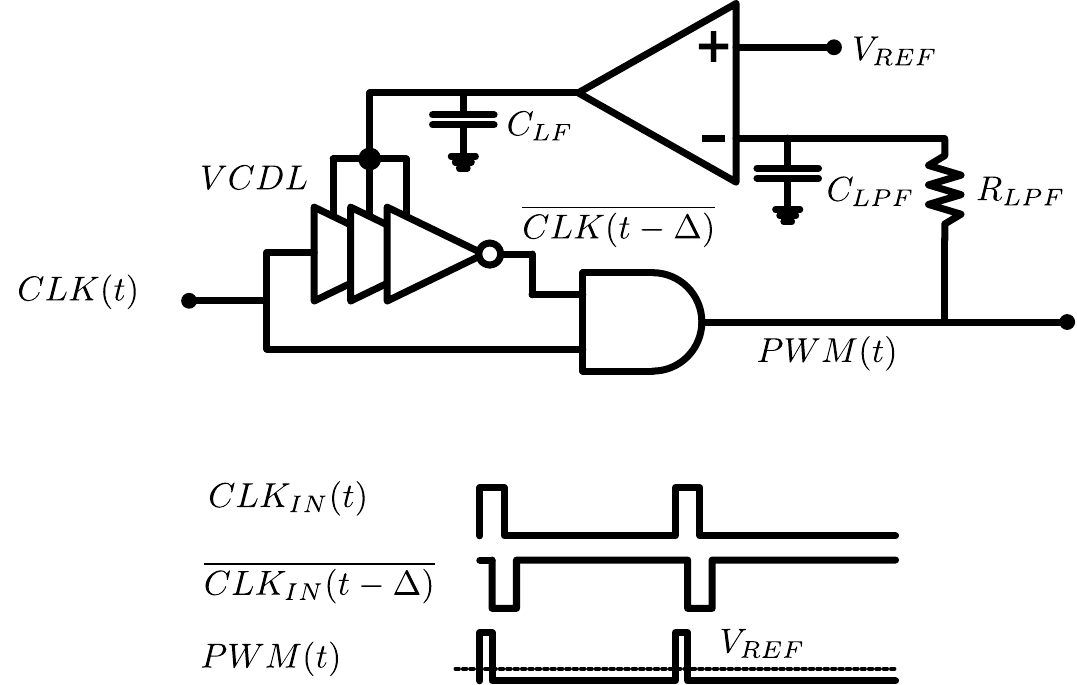}\caption{Generation of pulses with variable duty-cycle employing a DLL \label{fig:Generation-of-pulses}}
\end{figure}

An alternative approach  for generating high-frequency PWM clocks
is described in \cite{hkang18} (Fig. \ref{fig:Generation-of-pulses}).
The approach  uses a voltage-controlled delay line (VCDL) within a
delay-locked loop (DLL), and can be employed here. An input clock
$CLK(t)$ with periodicity of $f_{LO}$ and pulse width of $1/(2f_{PWM})$
is applied to the VCDL which provides delayed version $CLK(t-\Delta)$
at its output. The delayed clock is combined with the input to generate
a pulse train $PWM(t)$, whose duty cycle is such that its average
value is made equal to a pre-determined reference $V_{REF}$. For
generating a PWM signal with 16 pulses, such as that in Fig. \ref{fig:Generation-of-symmetrically},
16 such pulse generators, with 16 corresponding $V_{REF}$s are employed.
The outputs of the individual pulse generators are time-interleaved
in steps of $T_{PWM}$ and combined through a simple logic operation,
to generate a PWM signal corresponding to a sinusoidal LO \cite{hkang18}.

The VCDL is a cascade of inverters, while the OP-AMP in the loop needs
a very low gain-bandwidth product, since it is comparing a DC signal
only. As such this is a low-power, and yet accurate approach to generate
PWM pulses at GHz rates. PWM pulses that are equivalent to those generated
by comparison of sinusoids to trailing-edge and leading-edge ramps
can both be generated using this approach. This technique is mathematically
equivalent to the ramp-based approach.

\section{Simulation results \label{sec:Simulation-results}}

Simulations of the proposed architecture of Fig. \ref{fig:Generation-of-symmetrically}
are shown below. The basic principle is verified using ideal circuit
switches, with specified on-state resistance. The simulations assume
a series inductor of 35-nH although an inductor in the range of 20-40
nH is found to provide similar performance. A parasitic capacitance
of 100-fF is assumed at the input. The design is also simulated and
verified using switches implemented in a 65-nm commercial CMOS process,
which includes the full-device model. The design employs PWM with
frequency $f_{PWM}=1.6$ GHz and a sinusoidal LO with a fundamental
frequency $f_{LO}=100$ MHz. An ideal clocking network is employed
in the simulations, however it is expected that clocking design similar
to that previously demonstrated in silicon in \cite{hkang18} will
be applicable to this work.
\begin{figure}[tbh]
\centering\includegraphics[width=0.9\columnwidth]{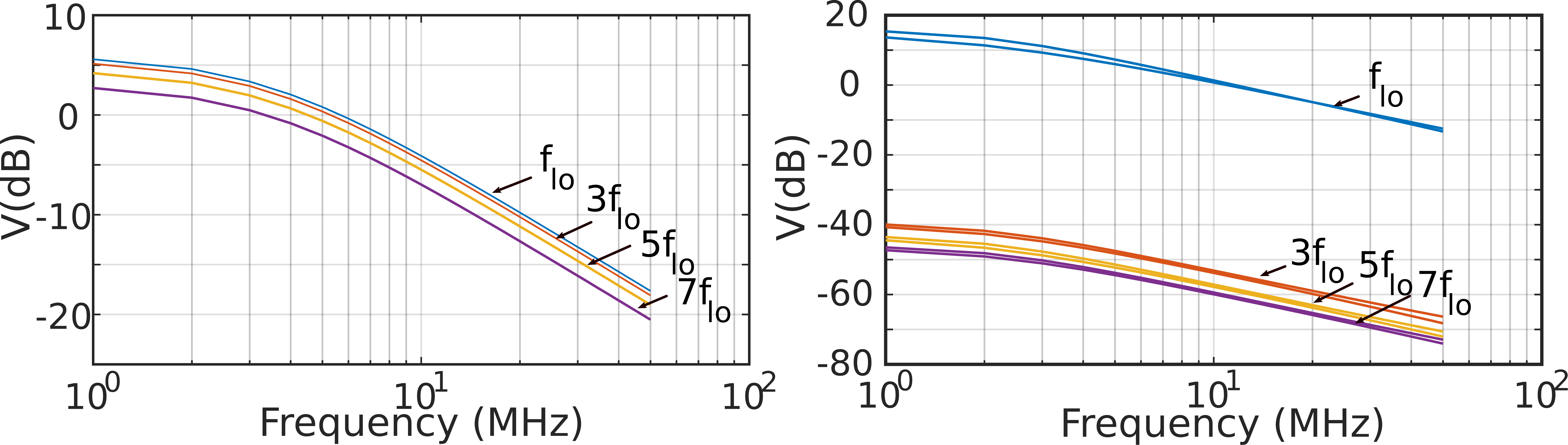}

\caption{Harmonic response comparison at baseband for 16-path filter with fixed
duty-cycle pulses and PWM-LO filter $f_{LO}=100$ MHz\label{fig:Harmonic-response-comparison}}
\end{figure}

The harmonic responses at baseband of an \textit{N}-path filter with
fixed $1/N$ duty-cycle for $N=16$, and PWM LOs in the configuration
of Fig. \ref{fig:Generation-of-symmetrically} are shown in Fig. \ref{fig:Harmonic-response-comparison}.
The input frequency is swept within a $50$ MHz band around $kf_{LO}$,
for $k=1,3,5,7$ and the frequency-translated output at baseband is
observed from around $kf_{LO}$. The harmonic-response is observed
to be significantly smaller for the PWM LOs.

\begin{figure}[tbh]
\centering\includegraphics[width=0.9\columnwidth]{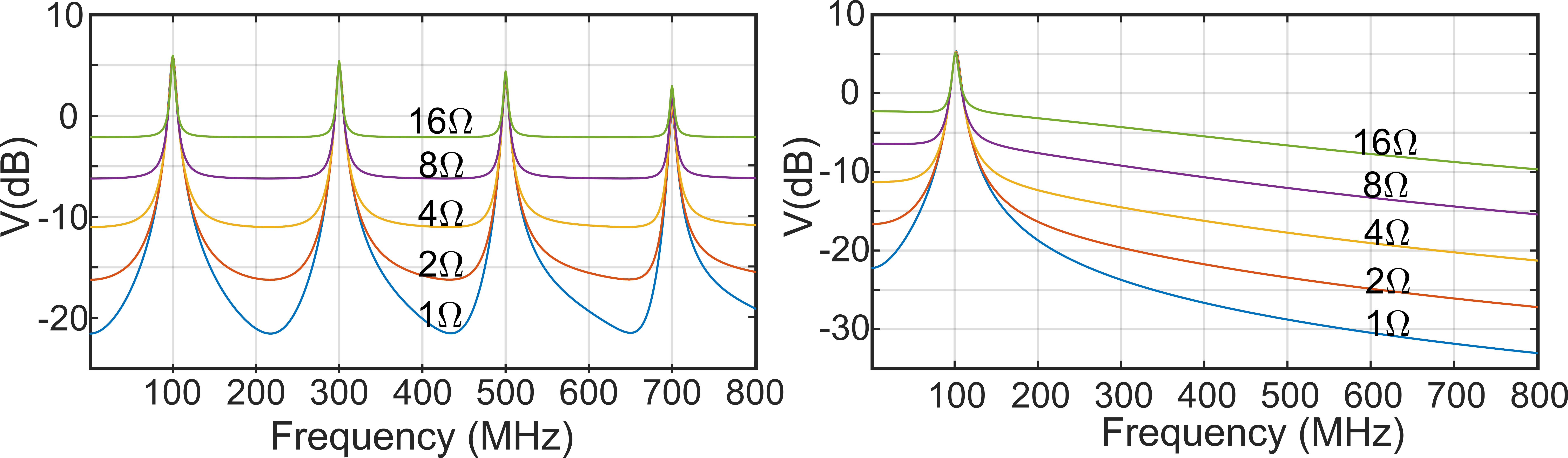}\caption{RF response for a 16-path filter with fixed duty-cycle pulses and
PWM-LO based design for different values of $R_{SW}$ ($f_{LO}=100$
MHz) \label{fig:RFinputwithswitchresistance}}
\end{figure}

The RF response for the above two cases is compared in Fig. \ref{fig:RFinputwithswitchresistance}
for various values of the switch resistance $R_{SW}$. The filter
with fixed duty-cycle LOs exhibits downconversion from all odd harmonics
of $f_{LO}$, while the PWM-LO based filter shows a response only
at $f_{LO}$. The response flattens with increasing $R_{SW}$ \cite{Ghaffari2010}.
The fixed duty-cycle filter simulations do not include an input inductor,
since it degrades the harmonic response.
 \vspace{-0.6cm}
\begin{figure}[tbh]
\centering\subfloat[RF input \label{fig:RFinput employing PWM-LOs}]{\centering\includegraphics[width=0.48\columnwidth]{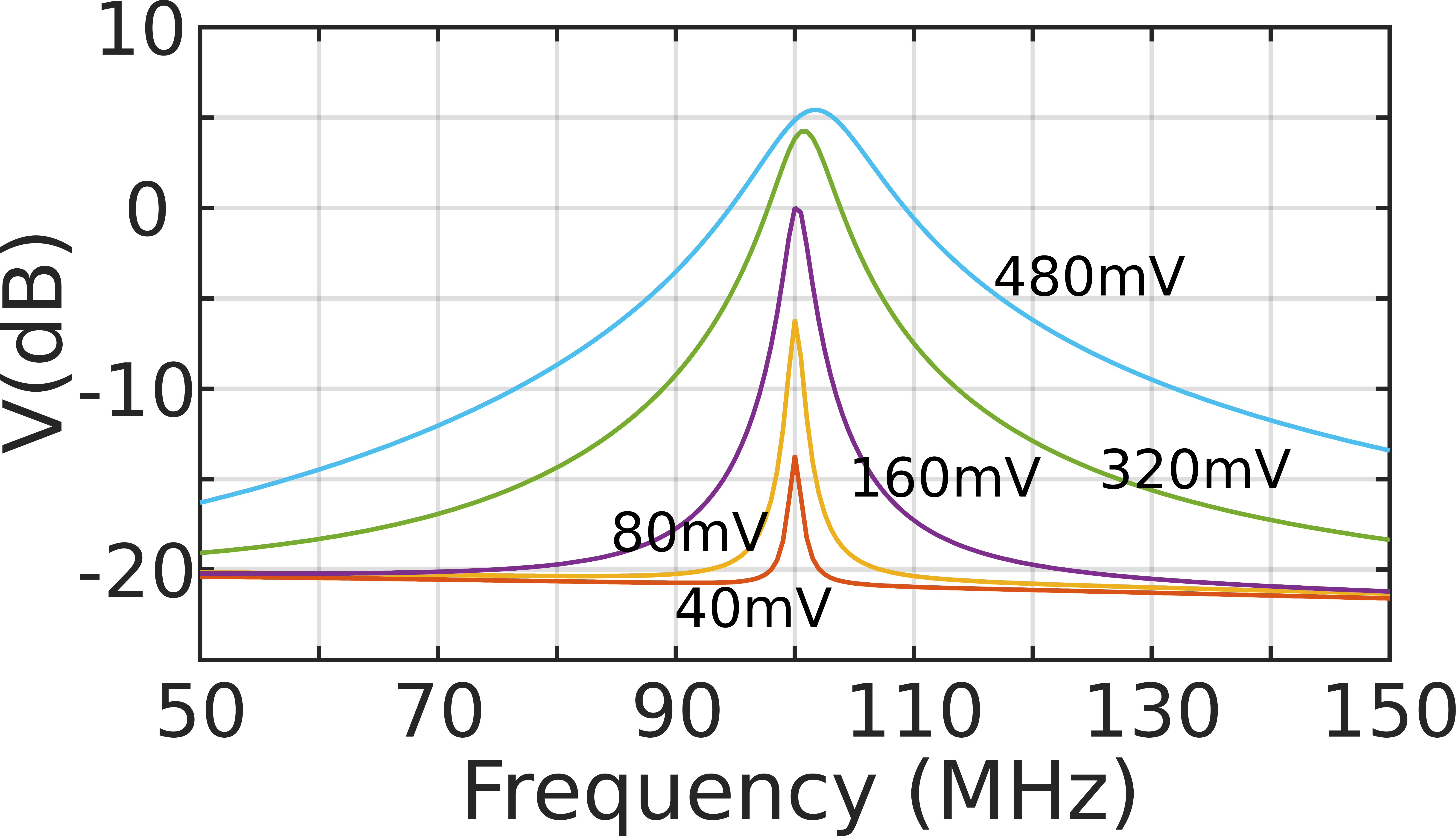}

} \subfloat[Baseband input \label{fig:BBoutput with PWM-LOs }]{\centering\includegraphics[width=0.48\columnwidth]{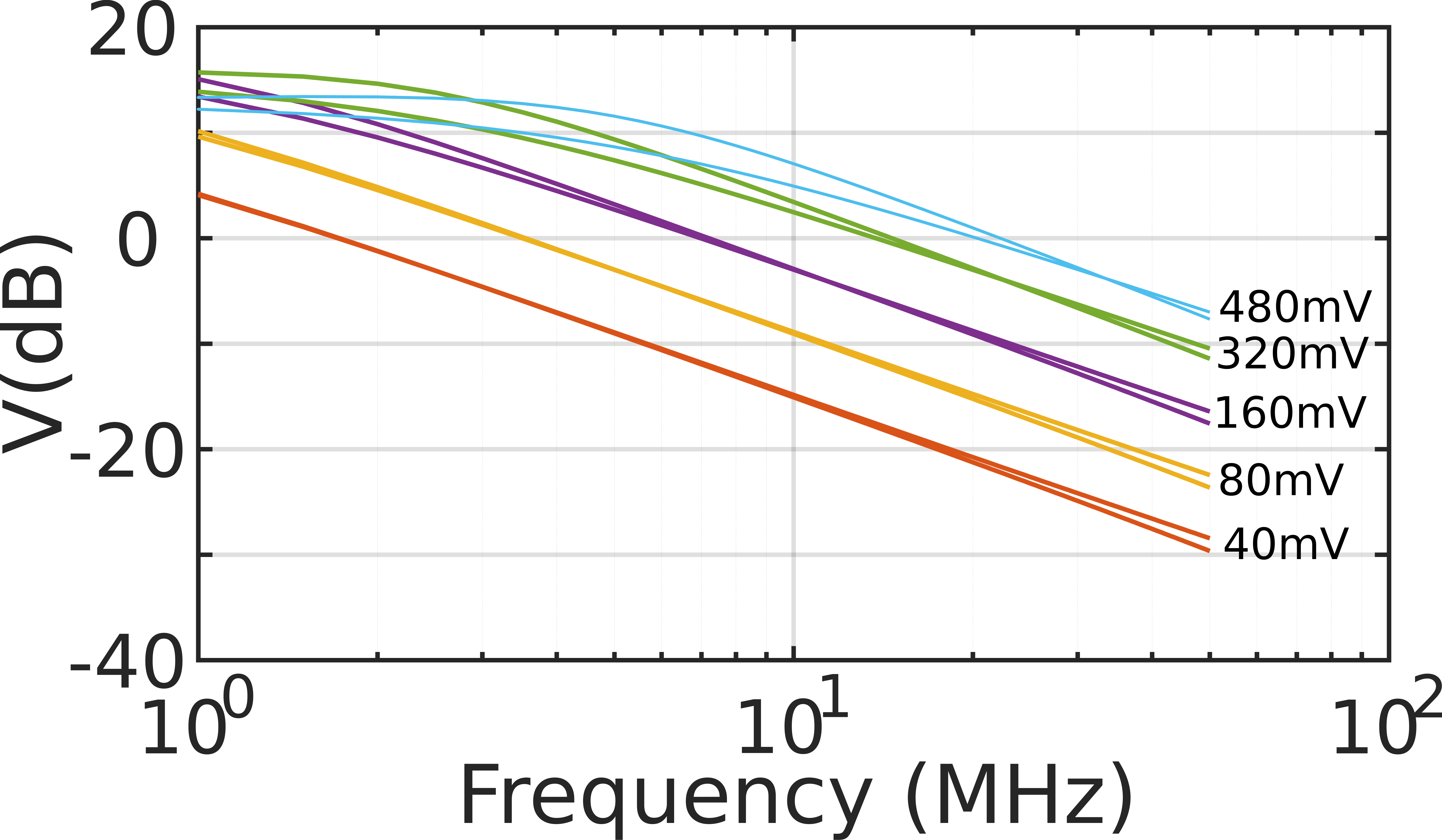}}\caption{Response for different sinusoidal LO amplitudes}
\end{figure}

The response of the $N$-path filter at the input node ($v_{rf}$
in Fig. \ref{fig:Generation-of-symmetrically}) for the desired RF
signal is shown in Fig. \ref{fig:RFinput employing PWM-LOs}, as a
function of the effective sinusoidal LO amplitude $A_{LO}$ (Fig.
\ref{fig:Generation-of-symmetrically}). A peak PWM ramp waveform
of 1.2 V, and a sinusoidal DC level of 0.6 V is assumed. The peak
of the bandpass response RF can be varied as a function of $A_{LO}$,
which implies that the impedance match at the input can be controlled
through $A_{LO}$. The maximum input voltage is observed for $A_{LO}\approx320$
mV, while an optimal input match is observed for $A_{LO}=160$ mV.
Increasing $A_{LO}$ to near the maximum value of $0.6$ V makes the
minimum pulse width close to zero, which is difficult to employ in
practice, due to rise and fall time limitations in the logic gates
and drivers. On the other hand, for $A_{LO}<300$ mV, the minimum
PWM pulse widths have a duty-cycle in excess of 20\%, which is achievable
in modern CMOS technologies for $f_{PWM}$ in the GHz range. The baseband
response ($v_{I}+-v_{I}-$ in Fig. \ref{fig:Generation-of-symmetrically})
is observed to vary linearly as LO amplitude (Fig. \ref{fig:BBoutput with PWM-LOs })
is increased and eventually saturate.

\begin{figure}[tbh]
\centering\includegraphics[width=0.47\columnwidth]{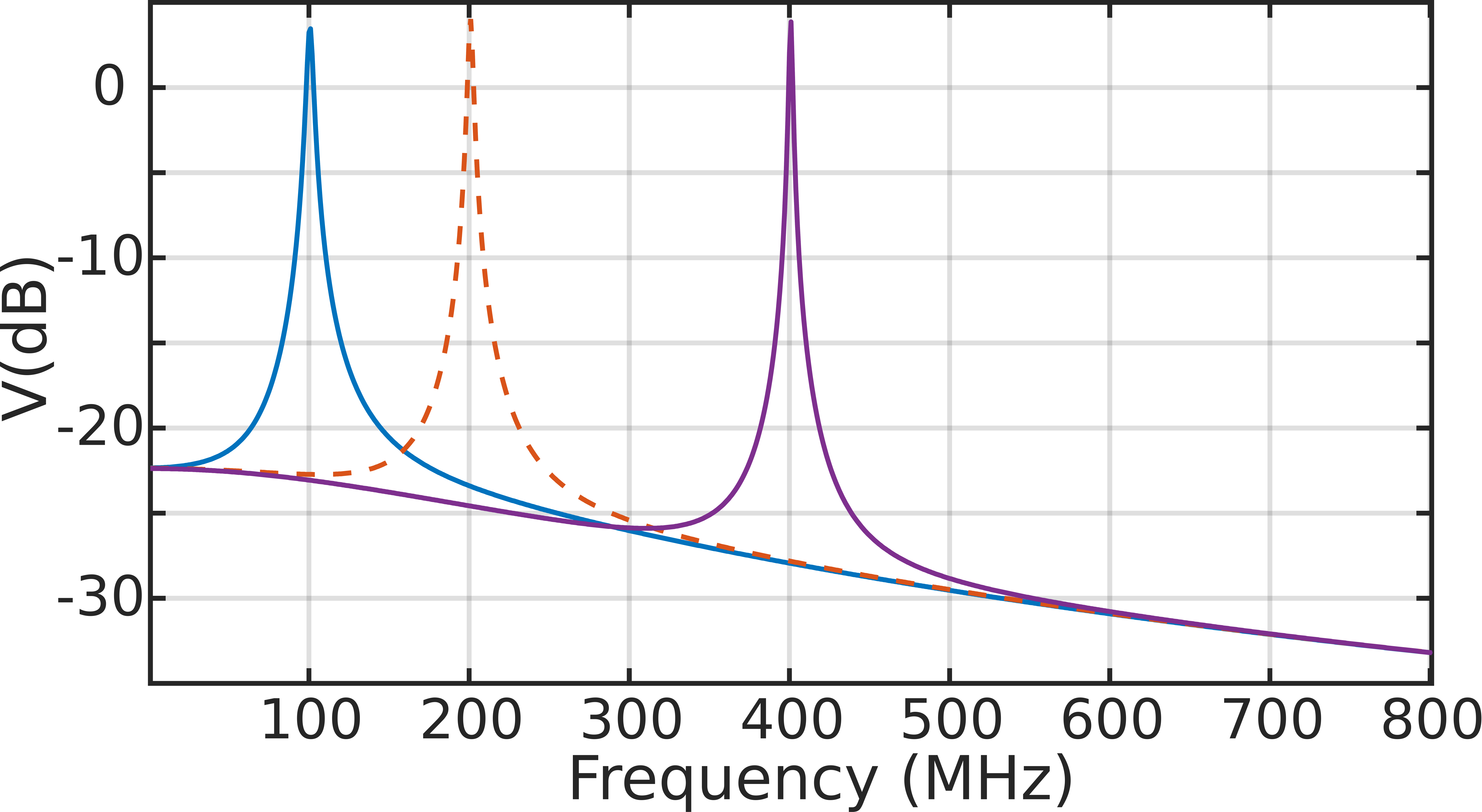}\caption{Fundamental response for $f_{LO}=100,200,400$ MHz\label{fig:Fundamental-response-for}}
\end{figure}

It is also possible to use sinusoids at multiples of $f_{LO}$ in
the PWM-LO filter, albeit at the expense of degraded harmonic folding.
The fundamental bandpass response for various $f_{LO}$s with $f_{PWM}=1.6$
GHz is shown in Fig. \ref{fig:Fundamental-response-for} at $k\times100$
MHz, for $k=1,$2,4. The corresponding worst-case in-band harmonic
folding are at levels of -65, -61 and -32 dB below peak, while considering
an input band of up to 800 MHz. These arise from the image response
of the filter and potentially can be reduced further.

\begin{figure}[tbh]
\centering\includegraphics[width=0.9\columnwidth]{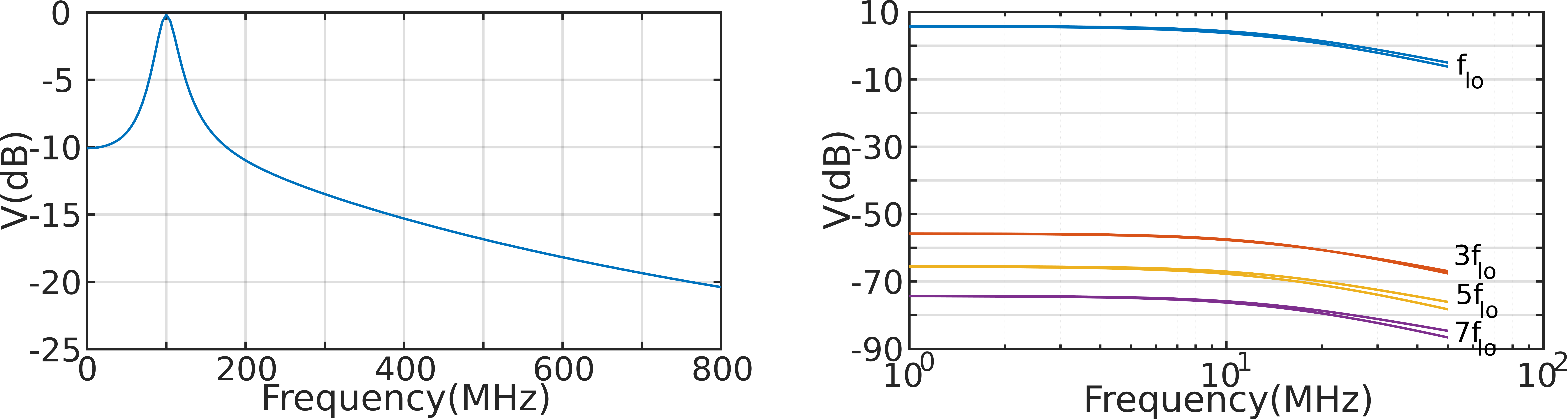}\caption{RF input and baseband response of PWM-LO  filter with MOSFETs \label{fig:Baseband-and-RF}}
\end{figure}

\begin{figure}[H]
\centering\includegraphics[width=0.9\columnwidth]{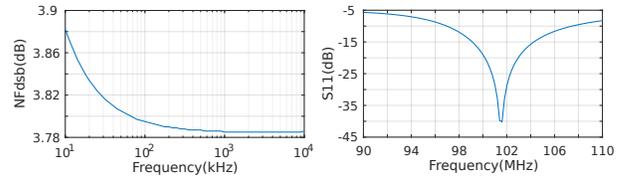}\caption{Simulated noise figure and S11 with MOS switches \label{fig:Simulated-noise-figure}}
\end{figure}

Simulation results employing MOS devices in a commercial 65nm CMOS
process are shown in Fig. \ref{fig:Baseband-and-RF}, which depicts
the RF and baseband response, including harmonic response. Similar
to using ideal switches, the design does not exhibit harmonic response.
Noise figure and S11 performance are shown in Fig. \ref{fig:Simulated-noise-figure}.

\vspace{-0.6cm}

\vspace{-0.6cm}

\section{Conclusion \label{sec:Conclusion}}

An $N$-path filter that employs multi-phase PWM clocks is proposed.
Each path of the $N$-path implements a sinusoidal LO, which effectively
avoid the harmonic responses observed in $N$-path implementations
that employ rectangular pulses. The design can allow for a a passive-mixer
based $N$-path implementation. Performance parameters such as gain
and bandwidth can be varied by changing the amplitude of the effective
LO sinusoid. Simulation results using ideal switches and physical
CMOS devices are presented.

\bibliographystyle{IEEEtran}
\bibliography{bibaddendumforetal,IEEEabrv,arXiv_ref}

\begin{thebibliography}{1}
\providecommand{\url}[1]{#1}
\csname url@samestyle\endcsname
\providecommand{\newblock}{\relax}
\providecommand{\bibinfo}[2]{#2}
\providecommand{\BIBentrySTDinterwordspacing}{\spaceskip=0pt\relax}
\providecommand{\BIBentryALTinterwordstretchfactor}{4}
\providecommand{\BIBentryALTinterwordspacing}{\spaceskip=\fontdimen2\font plus
\BIBentryALTinterwordstretchfactor\fontdimen3\font minus
  \fontdimen4\font\relax}
\providecommand{\BIBforeignlanguage}[2]{{%
\expandafter\ifx\csname l@#1\endcsname\relax
\typeout{** WARNING: IEEEtran.bst: No hyphenation pattern has been}%
\typeout{** loaded for the language `#1'. Using the pattern for}%
\typeout{** the default language instead.}%
\else
\language=\csname l@#1\endcsname
\fi
#2}}
\providecommand{\BIBdecl}{\relax}
\BIBdecl
\renewcommand{\BIBentryALTinterwordstretchfactor}{1}

\bibitem{Franks1960}
L.~E. {Franks} \emph{et~al.}, ``{A}n {A}lternative {A}pproach to the
  {R}ealization of {N}etwork {T}ransfer {F}unctions: {T}he {N}-{P}ath
  {F}ilter,'' \emph{{B}ell {S}ystem {T}echnical {J}ournal}, vol.~39, no.~5, pp.
  1321--1350, Sep. 1960.

\bibitem{Cook2006}
B.~W. {Cook} \emph{et~al.}, ``{Low-power 2.4-GHz transceiver with passive RX
  front-end and 400-mV supply},'' \emph{IEEE JSSC}, vol.~41, no.~12, pp.
  2757--2766, Dec. 2006.

\bibitem{Ghaffari2010}
A.~{Ghaffari} \emph{et~al.}, ``A differential 4-path highly linear widely
  tunable on-chip band-pass filter,'' in \emph{Proc. IEEE RFIC Symp.}, May
  2010, pp. 299--302.

\bibitem{Duipmans2015}
L.~{Duipmans} \emph{et~al.}, ``{A}nalysis of the {S}ignal {T}ransfer and
  {F}olding in {N}-{P}ath {F}ilters {W}ith a {S}eries {I}nductance,''
  \emph{IEEE TCAS-1}, vol.~62, no.~1, pp. 263--272, Jan. 2015.

\bibitem{Han2019}
G.~{Han} \emph{et~al.}, ``{A} 0.3-to-1.3{GH}z {M}ulti-{B}ranch {R}eceiver with
  {M}odulated {M}ixer {C}locks for {C}oncurrent {D}ual-{C}arrier {R}eception
  and {R}apid {C}ompressive-{S}ampling {S}pectrum {S}canning,'' in \emph{Proc.
  IEEE RFIC Symp.}, Jun. 2019, pp. 95--98.

\bibitem{Darvishi2013}
M.~{Darvishi} \emph{et~al.}, ``{D}esign of {A}ctive {N}-{P}ath {F}ilters,''
  \emph{IEEE JSSC}, vol.~48, no.~12, pp. 2962--2976, Dec. 2013.

\bibitem{xu18}
Y.~{Xu} \emph{et~al.}, ``A blocker-tolerant {RF} front end with
  harmonic-rejecting {$N$}-path filter,'' \emph{IEEE JSSC}, vol.~53, no.~2, pp.
  327--339, Feb 2018.

\bibitem{hkang18}
H.~{Kang} \emph{et~al.}, ``A wideband receiver employing pwm-based harmonic
  rejection downconversion,'' \emph{IEEE JSSC}, vol.~53, no.~5, pp. 1398--1410,
  May 2018.

\bibitem{Nielsen_PWM}
K.~Nielsens, ``{A review and comparison of pulsewidth modulation (PWM) methods
  for analog and digital input switching power amplifiers},'' in
  \emph{102$^{nd}$ {AES} {C}onvention}, March 1997.

\end{thebibliography}

\end{document}